\documentclass[prl,aps,showpacs,amsmath,twocolumn,floatfix]{revtex4}

\usepackage{epsfig}

\begin{document}

\title{Resonant thermal transport in semiconductor barrier structures}

\author{P.~Hyldgaard} 
\affiliation{Department of Applied Physics, Chalmers University of 
Technology, SE-41296 G{\"o}teborg, Sweden.} 
\date{January 18, 2004}

\begin{abstract} 
I report that thermal single-barrier (TSB) and thermal double-barrier 
(TDB) structures (formed, for example, by inserting one or two regions
of a few Ge monolayers in Si) provide both a 
suppression 
of the phonon transport as well as a resonant-thermal-transport 
effect. I show that high-frequency phonons can experience a traditional 
double-barrier resonant tunneling in the TDB structures while the 
formation of Fabry-Perot resonances (at lower frequencies)
causes quantum oscillations in the temperature variation
of both the TSB and TDB thermal conductances 
$\sigma_{\text{TSB}}$ and $\sigma_{\text{TDB}}$.
\end{abstract}


\maketitle

The understanding of phonon transport in nanoscale 
heterostructured materials~\cite{NanoThermal} is in an 
exciting development motivated in part by the 
search to improve both thermo-electric~\cite{HetTheDev} 
and thermo-ionic~\cite{ThermIon} cooling. 
The interest also derives from the observation
that nanostructure phonons exhibit nanoscale-transport,
confinement, and quantization effects similar to those 
observed for electrons and photons.  A significant 
suppression is observed~\cite{Yao,ChenAPL} in the in-plane 
thermal conductivity of heterostructures and is explained 
by interface scattering as a phonon Knudsen-flow 
effect~\cite{Knudsen,ChenTherm}. 
Similarly, the perpendicular thermal conductivity $\kappa_{\rm SL}$ 
of semiconductor superlattices shows a dramatic 
reduction~\cite{ChenHeat,Cahill} (compared to the average bulk 
conductivities) that cannot be accounted for alone by the
expected decrease~\cite{Dow} in the effective superlattice-phonon 
lifetime $\tau_{\rm SL}$. Instead, the strong reduction in 
$\kappa_{\rm SL}/ \tau_{\rm SL}$ results from a pronounced
miniband formation where the  difference in materials hardness 
forces an increasing confinement of modes with a finite
in-plane momentum to either the Si or Ge layers in the
superlattice~\cite{SupLat}. Finally, the phonon quantum-point-contact 
effect in nanoscale dielectric wires~\cite{phonQPC} shows that the 
phonon wave nature also directly affects the low-temperature phonon 
transport.

Here I extend the search for quantized thermal-transport 
effects in semiconductor nanostructures to finite temperatures.
I focus on the phonon conduction across 
Si\-/few-Ge-monolayers\-/Si thermal single-barrier (TSB) and 
corresponding Si/Ge/Si/Ge/Si thermal double-barrier (TDB) 
heterostructures. I document (i) a strong
suppression of the phonon-transport thermal conductances 
$\sigma_{\rm TSB}$ and $\sigma_{\rm TDB}$, 
(ii) a traditional type of 
double-barrier resonant tunneling for high-frequency phonons 
in the TDB structures, and (iii) that phonon Fabry-Perot 
resonances~\cite{FabryPerot} (at lower frequencies) produce 
a resonant-thermal-transport effect at finite temperatures 
in both the TSB and TDB structures (for sufficiently small 
Ge-barrier thicknesses).  My focus is on the technology relevance 
rather than the low-temperature transport. Instead of the 
long-wavelength phonon analysis of Ref.~\cite{phonQPC}, I 
therefore present a phonon-model calculation of the (resonant) thermal 
transport which remains applicable at finite frequencies, includes 
the phase-space limitations imposed by total-internal reflection, and 
describes the important effect of increasing misalignment of the 
Si-/Ge-phonon dynamics as the in-plane momentum $q$ 
increases~\cite{SupLat}. 
The prediction (iii) bridges the concept of 
thermal-transport quantization effects from the previous 
focus on low-temperature phonon 
transmission~\cite{phonQPC} to the finite-temperature 
nanostructure heat conduction for which I show that oscillations 
persist up to temperatures $\Theta\approx \hbox{50 K}$.

\begin{figure}
\centerline{\epsfig{figure=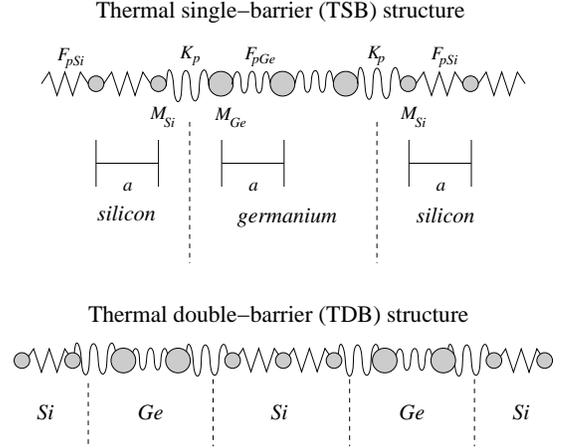,width=.4\textwidth}}
\caption{\label{fig1}
Sample realization of thermal barrier structures in which even 
a few monolayers of a softer material (here germanium) serve to 
inhibit the otherwise effective (phonon) thermal transport 
in a surrounding hard materials (here Si). The figure shows 
examples of both thermal single-barrier (TSB) and 
thermal double-barrier (TDB) heterostructures that 
exhibit a resonant-thermal transport effect. The figure also 
shows schematics of the phonon model 
used to
calculate the finite-temperature variation of the TSB 
and TDB thermal conductances. 
}
\end{figure}

To emphasize the potential technological relevance I first report
a simple estimate for the 
thermal-conductance 
suppression in two examples of TSB and TDB structures, Fig.~1, 
formed as a Si/triple-Ge-monolayer/Si and as a Si/2Ge/3Si/2Ge/Si 
semiconductor heterostructure, respectively.  I predict below 
a strong high-temperature suppression
\begin{equation}
\sigma_{\text{TSB,TDB}}\sim \sigma_{K} \lesssim 
\hbox{10$^{4}$ W/Kcm$^2$} 
\label{ApproxSigma}
\end{equation}
specified by the single-Si/Ge interface Kapitza
conductance~\cite{Kapitza} 
$\sigma_{\rm K}(\Theta)\leq \sigma_{\rm K}(\Theta\to\infty)\approx
\hbox{10$^{4}$ W/Kcm$^2$}$ (as also obtained in my phonon-model 
calculation).  The thermal-conductance estimate~(\ref{ApproxSigma}) 
results by describing the full dynamics of phonons that approach an 
individual single-barrier/double-barrier structure (rather than
moving in a superlattice) and therefore provides a more consistent 
description than the co-called effective conductance~\cite{Cahill} 
$G_{\rm SL} \equiv \kappa_{\rm SL}(d) /(d/2) \sim 10^{5} \hbox{W/Kcm$^2$}$ 
extracted~\cite{Cahill} from measurements at ($\Theta=\hbox{200 K}$)
of the thermal conductivity $\kappa_{\rm SL}(d)$ in Si/Ge superlattices
with a nanoscale periodicity $d$. The estimate~(\ref{ApproxSigma}) shows 
that repeating the TSB or TDB formation every 50 nm provides a 
reduction of the effective thermal conductivity $\kappa$ to the 
value of a Si/Ge alloy $\kappa_{\rm alloy} \approx \hbox{5 W/Km}$ 
(repetition every 5 nm would be needed if the limit $\sigma_{\rm TSB} 
\lesssim G_{\rm SL}$ applies). 

The TSB and TDB structures also give rise to a
resonant-thermal-transport effect which is observable 
at finite temperatures ($\Theta \lesssim \hbox{50 K}$).  
Fig.~1 shows a schematics of the phonon model which I here 
solve to calculate the phonon tunneling and the resulting 
thermal conductance across both a single interface 
and across the repeated interfaces in the TSB and TDB structures. 
I assume a shared silicon (germanium) lattice constant $a$, the 
atomic masses $M_{\text{Si(Ge)}}$, and intra-silicon
(intra-germanium) force constants $F_{p;\text{Si(Ge)}}$ and 
inter-layer coupling constant $K_p\equiv ({F_{p;\text{Si}} 
F_{p;\text{Ge}}})^{1/2}$ specified by the materials sound 
velocities~\cite{SupLat}.  The 
tunneling~\cite{Narayamurti} of phonon modes that are 
polarized and propagating in the perpendicular $\hat{z}$-direction 
is well described within such a one-dimensional lattice model. I
refer to our previous investigation of superlattice 
thermal transport~\cite{SupLat} for a description of how I 
include the effects of a finite in-plane momentum within
a simple-cubic model by adding in-plane force constants 
$F_{t;\text{Si(Ge)}}$ and corresponding characteristic frequencies 
$\Omega_{p,t;\text{Si(Ge)}} \equiv ({4 F_{p,t;\text{Si(Ge)}} 
/M_{\text{Si(Ge)}}})^{1/2}$; Fig.~1 of Ref.~\cite{SupLat}.  The set of 
force constants $F_{p,t;\text{Si(Ge)}}$ also specifies the 
phonon-transport contributions from $\xi_{\hat{x},\hat{y}}||
\hat{x},\hat{y}$-polarized heterostructure phonons~\cite{FitChoice}
and these modes are, of course, also included in the transport
calculation.  Below I limit the formal discussion to the 
contributions from $\xi_p|| \hat{z}$-polarized modes which
for a given in-plane momentum $\vec{q}$ can be characterized 
by the dimensionless in-plane energy measure~\cite{SupLat} 
$0< \alpha_{\vec{q}} \equiv 
\left[2 - \cos(q_x a) -\cos(q_y a)\right]<4$. 
This measure is, like the frequency, 
conserved~\cite{SupLat}
across the heterostructure interfaces
in this phonon-transport model study.

\begin{figure}
\centerline{\epsfig{figure=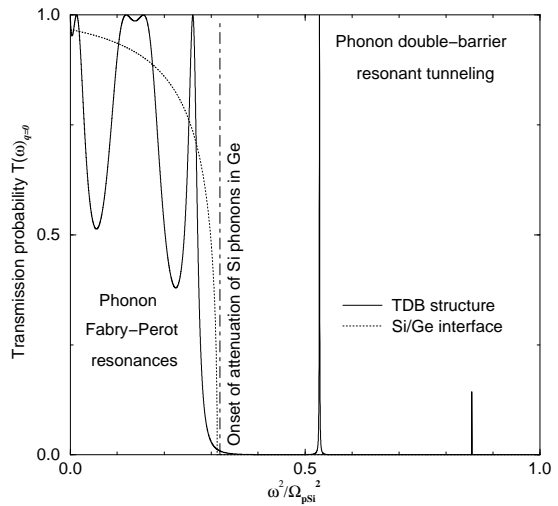,width=.4\textwidth}}
\caption{\label{fig2}
Comparison of calculated phonon transmission probability (evaluated at 
in-plane momentum $q=0$) for the Si/Ge thermal double-barrier 
structure, Fig.~1 (solid curve) and for the individual Si/Ge
interface (dotted curve). The vertical line identifies the
frequency-squared value above which incoming Si phonons become
attenuated in Ge. The TDB structure is seen to support 
a traditional type of double-barrier resonant tunneling above
this limit (when the Ge-layers represent actual barriers) but 
also multiple Fabry-Perot resonances (at lower frequencies)
when incoming phonons experience a partial transmission at
each of the individual Si/Ge interfaces. Note that the TDBs 
(and TSBs) naturally become completely transparent as $\omega\to0$.
}
\end{figure}

Figure 2 reports my calculations of the phonon transmission
in the single-interface, TSB, and TDB structures. Starting from 
the model equations of motion for every atom and adapting the
approach of Ref.~\cite{Kapitza}, I determine the 
single-interface, TSB, and TDB transmission probabilities 
$T_{\rm K,TSB,TDB}(\omega,\alpha_{\vec{q}})$ as a function 
of the phonon frequency and (conserved) in-plane momentum 
$q$. The figure reports a comparison of $T_{\rm K}
(\omega,\alpha_{\vec{q}=0})$ and $T_{\rm TDB}(\omega,
\alpha_{\vec{q}=0})$ and identifies the onset (vertical 
dashed-dotted line) when incoming Si phonons become
attenuated in the Ge barriers. The figure
documents that the TDB structure supports a traditional type 
of phonon (double-barrier) resonant tunneling at 
high frequencies ($\omega_>$) when the phonon dynamics in Ge is 
attenuated [and $T_{\rm K}(\omega_>)=0$]. 
The figure also illustrates that both the TDB and TSB 
structures support a number of Fabry-Perot resonances at
lower frequencies ($\omega_<$) when the individual interfaces 
provide partial transmission [i.e., when $0< T_{\rm K}(\omega_<)<1$].

To clarify the nature and measurability of the 
high-frequency resonant tunneling, 
I summarize the model calculations of the phonon dynamics 
at general in-plane momentum, $\alpha_{\vec{q}}\neq 0$.
I note that bulk silicon (germanium) phonon 
propagation at a given $\alpha_{\vec{q}}$ requires that
the frequency square exceeds $\omega^2_{\text{Si(Ge),min}} 
(\alpha_{\vec{q}}) \equiv \Omega^2_{t;\text{Si(Ge)}} 
(\alpha_{\vec{q}}/2)$ but remains bounded by 
$\omega^2_{\text{Si(Ge),max}} (\alpha_{\vec{q}})
\equiv \Omega^2_{t;\text{Si(Ge)}}(\alpha_{\vec{q}}/2)+ 
\Omega^2_{p;\text{Si(Ge)}}$~\cite{SupLat}. The finite 
silicon/germanium acoustic mismatch ensures that 
$\omega_{\text{Si,min[max]}} (\alpha_{\vec{q}}) > 
\omega_{\text{Ge,min[max]}}(\alpha_{\vec{q}})$. 
Phonon propagation (i.e., absence of attenuation)
in {\it both\/} silicon and germanium layers thus 
effectively requires that
\begin{equation}
\omega^2_{\text{Si,min}}(\alpha_{\vec{q}}) < \omega^2 <
\omega^2_{\text{Ge,max}}(\alpha_{\vec{q}}), 
\label{Om2SIGeCon}
\end{equation}
since this is the condition for an incoming silicon phonon to 
avoid total internal reflection at an individual Si/Ge interface
(as formulated in the present model study). The
condition~(\ref{Om2SIGeCon}) is a severe phase-space restriction which,
for example, becomes {\it impossible\/} to satisfy at $\alpha_{\vec{q}} > 
2$ in Si/Ge structures~\cite{SupLat}. For incoming Si phonons 
with a frequency $\omega$ above the onset of Ge attenuation,
the model study yields a strong exponential decay $1/\gamma_{\text{Ge}} 
\sim a$, a vanishing single-barrier transmission ($T_{\rm TSB}\to 0$),
and the strongly peaked double-barrier resonant-tunneling transmission 
$T_{\rm TSB}$ that is illustrated in Fig.~2. The high-frequency
resonant tunneling may become observable~\cite{Narayamurti} if 
it is possible to design a frequency-selective source of 
THz phonons~\cite{PhonSource}.
However, for a thermal distribution $N(\omega,\Theta)$ of incoming
Si phonons at a relevant elevated temperature $\Theta$, I find 
the phase-space contribution from the (high-frequency) resonant 
tunneling, Fig.~2, too small to directly affect the TDB thermal 
conductance.

In contrast, I find that the phonon Fabry-Perot resonances do 
produce observable finite-temperature quantization effects
in the TDB and TSB heat conduction, Fig.~3.  From the calculated 
phonon transmission probabilities I determine (adapting 
Ref.~\cite{Kapitza}) the single-Si/Ge interface and 
thermal-barrier conductances
\begin{eqnarray}
\sigma_{\text{K,TSB,TDB}} = & &
\sum_{m} \int\frac{d^2q}{(2\pi)^2} 
\times\nonumber \\
\Big[ 
\int_{\omega_{\text{min}}}^{\omega_{\text{max}}} 
\frac{d\omega}{2\pi}&& \hbar \omega \, 
T_{\text{K,TSB,TDB}}(\omega;\alpha_{\vec{q}})
\left(\frac{dN}{d\Theta}\right)\Big]_{m}
\label{SigDef}
\end{eqnarray}
by summing up the contributions at different modes
$m=\xi_p,\xi_{\hat{x},\hat{y}}$ and in-plane momenta $q$.
In the result~(\ref{SigDef}) $\omega_{\text{min}}$ and 
$\omega_{\text{max}}$ represent a shorthand for
the corresponding frequency-integration limits
$\omega_{\text{Si,min}}(\alpha_{\vec{q}})$ and
$\omega_{\text{Si(Ge),max}}(\alpha_{\vec{q}})$. I
stress that the condition~(\ref{Om2SIGeCon}) must be 
absolutely satisfied to obtain a transport contribution to 
the interface conductance $\sigma_{\text{K}}$ and 
effectively satisfied for the TSB and TDB conductances 
$\sigma_{\rm TSB,TDB}$ (however, all TSB and TDB-transport
contributions are retained in the calculations reported here).

\begin{figure}
\centerline{\epsfig{figure=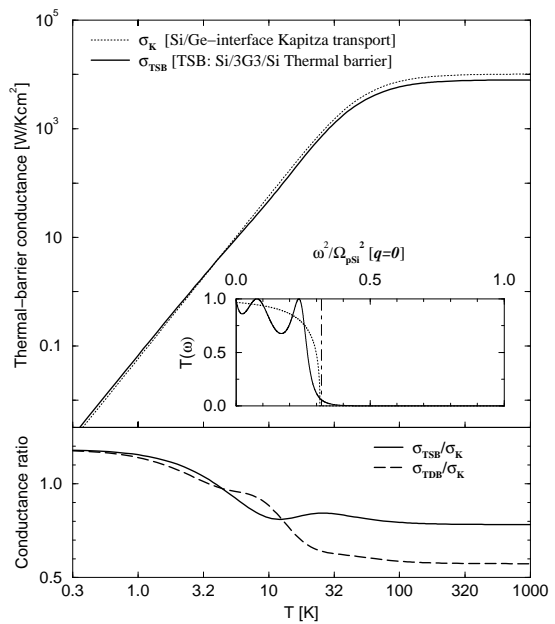,width=.4\textwidth}}
\caption{\label{fig3}
Temperature dependence of (perpendicular) thermal conductance
$\sigma_{\text{TSB}}$ (solid curve) for a Si/3Ge/Si TSB
heterostructure compared with the Si/Ge-interface Kapitza
conductance $\sigma_{K}\equiv \sigma_{\text{Si/Ge}}$ (dotted curve).
The thermal barrier transport is effectively subject to similar
severe phase-space restrictions (enforced by total internal 
reflection) as the single-interface conductance and can thus be
estimated  $\sigma_{\text{TSB}}(T)\sim \sigma_{\text{K}}(T)$. 
{\it The insert panel\/} compares the TSB phonon transmission 
probability $T_{\text{TSB}}(\omega)\equiv T_{\text{TSB}}
(\omega;\alpha_q=0)$ (solid curve) and the single-Si/Ge transmission
probability (dotted curve) at zero in-plane momentum, $q=0$.  The barrier 
transmission is seen to be essentially eliminated above the frequency 
(vertical dashed-dotted line) when the incoming phonon must tunnel across 
the Ge barrier layers. At the same time, the barrier transmission $T_{\rm
TSB}$ also shows clearly defined low-energy phonon Fabry-Perot 
resonances (similar to those of the $T_{\rm TDB}$, Fig.~2) produced 
within the 3Ge barrier.  {\it The bottom panel\/} documents 
how the set of lower-energy Fabry-Perot resonances produce quantum 
oscillations in the TSB and TDB conductance ratios $\sigma_{\text{TSB}}/
\sigma_{\text{K}}$ (solid curve) and $\sigma_{\text{TDB}}/
\sigma_{\text{K}}$ (dashed curve).
}
\end{figure}

The top panel of Fig.~3 illustrates the general validity 
of the ``classical'' thermal-conductance approximation, 
Eq.~(\ref{ApproxSigma}), and documents the additional 
temperature variation produced by the low-energy Fabry-Perot 
resonances. The panel shows that the thermal-barrier conductance 
$\sigma_{\text{TSB}}$ (solid curve) at all temperatures is 
comparable to and generally smaller than the individual Si/Ge-interface 
conductance $\sigma_{\text{K}}$ (dotted curve). 

The insert panel compares the corresponding phonon 
transmission probability $T_{\text{TSB}} (\omega)$ 
(calculated for Si/3Ge/Si at $\vec{q}=0$) and the 
single-Si/Ge-interface transmission (dotted curve).
The insert motivates the estimate~(\ref{ApproxSigma}) by 
emphasizing that (a) the thermal single-barrier transport 
is effectively restricted to incoming Si modes that satisfy the
condition~(\ref{Om2SIGeCon}) (exactly as was found for the TDB 
transport, Fig.~2) and (b) the single-interface transmission 
(dotted curve) approximates the {\it average\/} thermal-barrier 
transmission 
\begin{equation}
\langle T_{\text{TSB}}(\omega)\rangle\sim \langle T_{\text{K}}(\omega)\rangle
\approx T_{\text{Si/Ge}}^0 \equiv \frac{4 Z_{\text{Si}}Z_{\text{Ge}}}
{(Z_{\text{Si}}+Z_{\text{Ge}})^2},
\label{ShoeString}
\end{equation}
specified by the differences in acoustic impedances,
$Z_{\text{Ge}}/Z_{\text{Si}} \approx 1.5$~\cite{Narayamurti}:
$T^0_{\text{Si/Ge}}\approx 0.95$. The observations (a) and (b)
complete the argument (proof) for the ``classical'' 
thermal-conductance estimate~(\ref{ApproxSigma}).

\begin{figure}
\centerline{\epsfig{figure=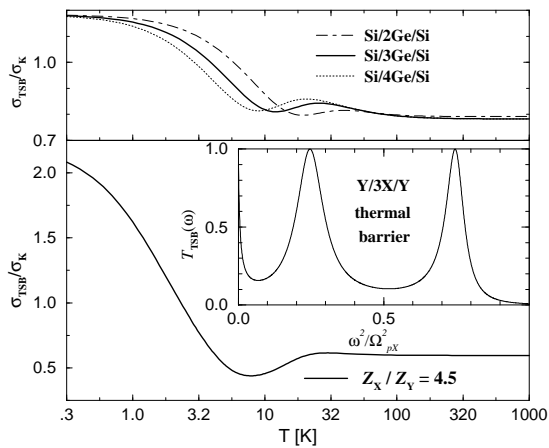,width=.4\textwidth}}
\caption{\label{fig4}
Robustness (top panel) and enhancement (bottom panel)
of the resonant-thermal-transport effects. 
The figure reports
calculations of the thermal conductance
ratio $\sigma_{\rm TSB}/\sigma_{\rm K}$ for both 
a set of Si/Ge/Si systems with different numbers of layers 
in the Ge barrier, and 
a fictitious X/3-Y-monolayer/X TSB structure with the significantly larger 
acoustic-impedance ratio $Z_{X}/Z_{Y} \approx 4.5$~\protect\cite{MgHf}.  
{\it The top panel\/} documents that the predicted 
TSB resonant-thermal-transport effect is robust as the 
set of different barrier thicknesses produce qualitatively 
identical oscillations in the temperature variation of the 
TSB thermal-conductance ratio.
{\it The insert and bottom panels\/} 
show 
that increasing the ratio of acoustic 
impedances causes much stronger Fabry-Perot resonances in the 
TSB transmission and an amplification of the thermal-conductance 
oscillations with temperature.
}
\end{figure}

The bottom panel of Fig.~3 documents that the phonon Fabry-Perot 
resonances refine the classical estimate~(\ref{ApproxSigma}) in
producing resonant-thermal-transport effects. These arise at higher 
temperatures and in a more general and technologically relevant 
group of structures than the previously investigated case of 
dielectric quantum wires, Ref.~\cite{phonQPC}.  Specifically, 
the bottom panel of Fig.~3 details quantum oscillations in the 
temperature variations for both $\sigma_{\text{TSB}}/ 
\sigma_{\text{K}}$ (solid curve) and 
$\sigma_{\text{TDB}}/ \sigma_{\text{K}}$ (dashed curve).
In both structures the thermal-conductance oscillations 
enhance as $T\to 0$ where the 
$T_{\rm TSB}(\omega)$ and $T_{\rm TDB}(\omega)$ 
variation increases in relative importance.

Finally, Fig.~4 emphasizes that the resonant-thermal-transport effects 
are robust against variations in the barrier thickness and
enhance with an increasing ratio of the acoustic impedances 
of the TSB and TDB heterostructures. The figure reports 
calculations of the thermal-conductance variation 
both for a set of Si/Ge/Si systems and 
in a fictitious TSB 
structure~\cite{MgHf} X/3Y/X where $Z_{X}/Z_{Y} \approx 4.5$.
The latter change produces a barrier transmission (insert
panel) where 
the Fabry-Perot resonances cause a much more dramatic deviation 
from the average transmission approximated
by the long-wavelength single-interface estimate~(\ref{ShoeString}),
$\langle T_{\rm K}(\omega) \rangle \sim 0.6$. 

{\it In summary,} I have investigated the phonon transport perpendicular 
to the interfaces of (silicon/triple-germanium-layer/silicon) 
thermal single-barrier (TSB) and corresponding thermal double-barrier 
(TDB) structures. I document a strong 
suppression of 
the finite-temperature heterostructure thermal conductances
$\sigma_{\text{TSB}}$ and $\sigma_{\text{TDB}}$ 
which approximately are limited by the
conductance $\sigma_{\rm K}$ of an individual Si/Ge interface. 
In addition, I predict quantum oscillations in the thermal-conductance 
ratios $\sigma_{\text{TSB}}/\sigma_{K}$ and $\sigma_{\text{TDB}}/
\sigma_{K}$ which arise from phonon Fabry-Perot resonances trapped 
in the central barrier or double-barrier region, respectively.

Discussions with G.~D.~Mahan are gratefully acknowledged.
This work was supported by the Swedish Foundation for Strategic
Research (SSF) through ATOMICS.


\begin{thebibliography}{99}
\bibitem{MgHf} The impedance ratio $Z_{\text{X}}/Z_{\text{Y}} 
\approx 4.5$ is very large but would characterize the phonon 
transport in a hypothetical Mg/triple-Hf-monolayer/Mg structure.
\end{thebibliography}
\end{document}